\documentstyle[preprint,aps,eqsecnum]{revtex}
\draft
\begin{document}
\bibliographystyle{plain}
\newcommand{\be}{\begin{equation}}
\newcommand{\ee}{\end{equation}}
\newcommand{\bea}{\begin{eqnarray}}
\newcommand{\eea}{\end{eqnarray}}

\title
{Symmetry Decomposition of Potentials with Channels}

\author{N. D. Whelan}
\address{Division de Physique Th\'{e}orique\cite{cnrsbullshit}, 
Institut de Physique Nucl\'{e}aire, 91406 Orsay Cedex, France.}

\date{\today}

\maketitle

\begin{abstract}
We discuss the symmetry decomposition of the average density of states for
the two dimensional potential $V=x^2y^2$ and its three dimensional 
generalisation $V=x^2y^2+y^2z^2+z^2x^2$. In both problems, the
energetically accessible phase space is non-compact due to the
existence of infinite channels along the axes. It is known that in
two dimensions the phase space volume is infinite in these
channels thus yielding non-standard forms for the average density of
states. Here we show that the channels also result in the symmetry
decomposition having a much stronger effect than in potentials without
channels, leading to terms which are essentially leading order.
We verify these results numerically and also observe a peculiar
numerical effect which we associate with the channels. In three
dimensions, the volume of phase space is finite and the symmetry
decomposition follows more closely that for generic potentials ---
however there are still non-generic effects related to some of
the group elements.
\end{abstract}

\section*{Introduction}

The role of chaos in the classical Yang-Mills fields has been 
examined by several authors, the studies typically being divided into two
r\'{e}gimes. In the first, one studies the full field theory \cite{fft} 
and tries to determine such global measures of chaos as the spectrum of
Lyapunov exponents and spatial-temporal correlations. In the second, one
studies the homogeneous or zero-dimensional limit of the problem
\cite{zerod} which admits a more microscopic analysis. Following this approach,
one is led to consider the three dimensional potential $V=x^2y^2+y^2z^2+z^2x^2$
and its simpler two dimensional cousin $V=x^2y^2$. The two dimensional
problem has been independently studied since it is an interesting
dynamical system in its own right. Until Dahlqvist and Russberg showed
otherwise \cite{dahlruss}, it was commonly believed that the classical
motion in this potential was completely chaotic. Although this is not
true, it remains one of the most chaotic potential systems known. It
also serves as a useful example of intermittency \cite{dahl1,dahl2}. Far from
the origin, the motion is confined within one of four channels within
which the problem is adiabatic so that a trajectory behaves in a smooth,
regular manner. Upon exiting the channel, the trajectory undergoes a
burst of strongly irregular motion before re-entering one of the
channels. This form of dynamics, regular behaviour with episodes of 
irregularity, is called intermittency and is found in
various physical systems including the classical helium atom
\cite{helium} and the hydrogen atom in a strong magnetic field 
\cite{hydrogen}. The first of these is governed by a potential very similar
in form to $x^2y^2$ \cite{eckwin}. The 
three dimensional problem shares the properties of strong chaos
and intermittency although this has been less extensively studied.

We will be interested in the requantisation of these potentials,
particularly in their densities of states. As proved by Simon
\cite{simon} and later analysed in greater detail by Tomsovic
\cite{tom}, the two dimensional potential has a discrete quantum
spectrum in spite of having an energetically accessible phase space of
infinite volume. This potential therefore violates
the semiclassical relation that the average number of
quantum energy levels below energy $E$ is proportional to the
volume of energetically accessible classical phase space.
In this paper we discuss a related property of this potential -
the manner in which the average density of states decomposes among the
various irreducible representations (irreps). Normally, the ratio of the 
number of states belonging to a given irrep $R$ 
of dimension $d_R$ is roughly $d_R^2/|G|$ \cite{pavloff,us}, 
where $|G|$ is the order of the group. There are then
small $\hbar$ corrections depending on the symmetry properties of the irreps
\cite{us}. We will show here that for the potentials mentioned above, the 
symmetry ``corrections'' can be anomalously large and in two
dimensions are essentially leading order in their effect.

The relevant symmetry groups for
the two and three dimensional potentials are $C_{4v}$ and the extended
octahedral group respectively (``extended'' because we allow for
inversions as well as rotations.) These groups have 5 and 10 conjugacy
classes of group elements, respectively, and we need to analyse them
all in order to calculate the average density of each irrep. The
method for doing this when there are no channels was discussed in 
Ref.~\cite{btu} for reflection operations and \cite{sw} in the context
of the permutation group of symmetric groups. It was then developed in
a more general context in Ref.~\cite{us}. For some of the classes
which appear here, the analysis is a straight-forward application of
this theory. For other classes, however, the channel
effects make it inapplicable and we use a different analysis
based on the adiabatic nature of the Hamiltonian deep in the channels
as introduced in \cite{tom}.  In both two and three dimensions, each
channel calculation involves an analysis of the subgroup
which leaves that channel invariant.

The structure of the paper is as follows. In the next section, we
review the formalism used in constructing the average density of states
from approximations of the heat kernels. The approximations are based
on Wigner transforms of the Hamiltonian and of unitary transformations
which correspond to the group elements. This formalism will be used in
the central region of the potential but will be adapted for
application to the channels. In section II we apply this to the two
dimensional potential and show that there are
very strong effects arising from this decomposition 
- much stronger than what one would expect for a normal bound potential. 
In section III, we verify these results numerically and also point out
the existence of a subtle numerical effect which is only apparent on
doing the symmetry decomposition. In section IV we introduce the three
dimensional generalisation and discuss the Wigner transforms
corresponding to the various group elements. In section V we do the
channel analysis of the three dimensional problem and use this to get
the final results for all classes. In three dimensions,
the channel effects are less dramatic but still introduce
modifications to what one expects for a generic potential.

\section{Formalism}

We will interest ourselves in the smooth average part $\bar{\rho}(E)$
of the density of states, often called the Thomas-Fermi term. There is
also an oscillating part $\rho_{\mbox{osc}}(E)$ given by periodic
orbits \cite{gutz} but we will not discuss this in great detail so in
what follows we suppress the bar on the smooth functions. The
specification of only concerning ourselves with the Thomas-Fermi
term in the density of states is made by invoking $\hbar$ expansions
rather than expansions involving oscillatory
functions of $1/\hbar$. One way to find the Thomas-Fermi density of
states is to work with the partition function (often called the heat
kernel), which is the Laplace transform of the density of states,
\be \label{hk}
Z(\beta) = \mbox{Tr}\left(e^{-\beta\hat{H}}\right) = {\cal
L}\left(\rho(E)\right).
\ee
In the presence of a symmetry group, each quantum state will belong to
one specific irreducible representation of that group so we will
consider the heat kernels of each irrep separately,
\be \label{phk}
Z_R(\beta) = \mbox{Tr}\left(\hat{P}_Re^{-\beta\hat{H}}\right).
\ee
$\hat{P}_R$ is the projection operator onto the irrep $R$ and
for a discrete group is given by \cite{hamermesh}
\be \label{proj}
\hat{P}_R = {d_R \over |G|}\sum_g\chi^*_R(g)\hat{U}(g).
\ee
The sum is over the elements of the group, $|G|$ in number,
$\chi_R(g)$ is the character of group element $g$ in irrep $R$, $d_R$
is the dimension of irrep $R$ and
$\hat{U}(g)$ is the unitary operator corresponding to the element $g$,
\be \label{unit}
\langle{\bf r}|\hat{U}(g)|\psi\rangle = \langle
g^{-1}{\bf r}|\psi\rangle = \psi(g^{-1}{\bf r}).
\ee

One standard way to proceed \cite{jbb} is to find the Wigner transform of the
operators $e^{-\beta\hat{H}}$ and $\hat{P}_R$ and integrate them to
evaluate the trace. The Wigner transform $A_W({\bf q,p})$ of a quantum
operator 
$\hat{A}$ is a representation of it in classical phase space and is defined as
\be \label{wt}
A_W({\bf q,p}) = \int d{\bf x} \left\langle{\bf q} + {{\bf x} \over
2}\right|\hat{A}\left|{\bf q} - {{\bf x} \over
2}\right\rangle e^{-i{\bf p\cdot x}/\hbar}.
\ee
To leading order in $\hbar$, it is valid to replace
$\left(e^{-\beta\hat{H}}\right)_W$ by $e^{-\beta H_W}$ where the
Wigner transform of the quantum Hamiltonian is just the
corresponding classical Hamiltonian. 
Traces are simply evaluated in this representation since
\bea
\mbox{Tr}(\hat{A}) 
& = & {1\over (2\pi\hbar)^n}\int d{\bf q}d{\bf p}
A_W({\bf q,p}) \nonumber \\
\mbox{Tr}(\hat{A}\hat{B}) & = & {1\over (2\pi\hbar)^n}\int d{\bf q}d{\bf p}
A_W({\bf q,p})B_W({\bf q,p}), \label{trwt}
\eea
where $n$ is the dimension of the system. As we will see below,
na\"{\i}ve application of these formulas may diverge in the channels; 
nevertheless, the formalism can be adapted.

In the evaluation of the Wigner transform of the projection operators
(\ref{proj}), we need the Wigner transforms of the unitary
operators $\hat{U}(g)$. This is discussed in detail in Ref.~\cite{us};
the results for all possible group elements in two dimensions are,
\bea
\left(\hat{U}(I)\right)_W({\bf q,p}) & = & 1 \nonumber \\
\left(\hat{U}(\sigma_i)\right)_W({\bf q,p}) & = &
\pi\hbar\delta(q_i)\delta(p_i) \nonumber \\
\left(\hat{U}(R_\theta)\right)_W({\bf q,p}) & \approx &
{\pi^2\hbar^2\over \sin^2({\theta\over 2})}
\delta(q_1)\delta(q_2)\delta(p_1)\delta(p_2).
\label{res}
\eea
The Wigner transform of the identity operator gives unity; the
transform of a reflection operator gives the delta functions of the
position and momentum corresponding to the symmetry plane; and, the
transform of a rotation gives the delta functions evaluated at the
symmetry axis. (The third result is exact for
$\theta=\pi$, otherwise it has higher order $\hbar$ contributions.)
An additional useful property is that the Wigner transform of the
product of two commuting operators is simply the product of their
respective transforms. Using this, we obtain from (\ref{res}) the
following relations for the three dimensional operators
\bea
\left(\hat{U}(\sigma_1\sigma_2\sigma_3)\right)_W({\bf q,p}) & = &
\pi^3\hbar^3 \delta({\bf q})\delta({\bf p})\nonumber\\
\left(\hat{U}(\sigma R_\theta)\right)_W({\bf q,p}) & \approx &
{\pi^3\hbar^3\over \sin^2({\theta\over 2})}
\delta({\bf q})\delta({\bf p}).
\label{3dres} 
\eea
The first of these says that the transform of the product of three
perpendicular reflections gives delta functions in all coordinates and momenta.
The second says that the transform of a reflection through a plane
times a rotation about the perpendicular axis gives the same delta functions.
In both (\ref{res}) and (\ref{3dres}), the relative power of
$\hbar$ equals the co-dimension of the set of points left invariant
by the group element.  We follow Ref.~\cite{us} in constructing
``class heat kernels'' 
\be \label{chk}
Z(g;\beta) = \mbox{Tr}\left(\hat{U}(g)e^{-\beta\hat{H}}\right)
\ee
so that
\be \label{dodo}
Z_R(\beta) = {d_R \over |G|} \sum_g\chi^*_R(g)Z(g;\beta).
\ee
The functions defined in Eq.~(\ref{chk}) are ``class
functions''; they do not depend explicitly on the group
element $g$ but only on the class to which it belongs.

\section{The potential $V=\lowercase{x}^2\lowercase{y}^2$}

The equipotential curves of this potential are shown as the light
curves in Fig.~\ref{system}. The symmetry group is $C_{4v}$, the same
as that of the square.  It consists of 8 elements: the identity
$\{I\}$; reflections through the channel axes $\{\sigma_x,\sigma_y\}$;
reflections through the diagonal axes $\{\sigma_1,\sigma_2\}$;
rotations by angle $\pi/2$, $\{R_{\pi/2},R_{-\pi/2}\}$; and, rotation
by angle $\pi$, $\{R_\pi\}$. These five sets of elements comprise the
five conjugacy classes. It follows that there are five irreps, four
are one dimensional and one is two dimensional. The character table is
given in Table 1.

We set out to calculate the five heat kernels corresponding to the
five classes. The integral corresponding to the identity is
\be \label{Id}
Z(I;\beta) = {1\over(2\pi\hbar)^2} \int dxdydp_xdp_ye^{-\beta H},
\ee
where $H=(p_x^2+p_y^2)/2 + x^2y^2$ is the classical Hamiltonian (and
the Wigner transform of the quantum Hamiltonian). This integral is
extensively discussed in Ref.~\cite{tom} where it is shown that it has
a logarithmic divergence. We return to this point below. According to
Eqs.~(\ref{trwt}) and (\ref{res}), the integral
corresponding to $\sigma_y$ is given by
\bea
Z(\sigma_y;\beta) & = & {1\over(2\pi\hbar)^2} \int
dxdydp_xdp_ye^{-\beta H}\pi\hbar\delta(y)\delta(p_y) \nonumber\\
& = & \sqrt{1\over{8\pi\beta\hbar^2}}\int dx. \label{refy}
\eea
Since the $x$ integral runs from $-\infty$ to $\infty$, this
integral diverges even more violently than (\ref{Id}). We will also
return to consider this more carefully below.
The remaining three class heat kernels are well behaved. For the
reflection through the diagonal axis
$\sigma_1$, we change variables to $\xi=(x+y)/\sqrt{2}$ and
$\eta=(x-y)/\sqrt{2}$ so that
\bea
Z(\sigma_1;\beta) & = & {1\over(2\pi\hbar)^2} \int
d\xi d\eta dp_\xi dp_\eta e^{-\beta
H}\pi\hbar\delta(\eta)\delta(p_\eta)
\nonumber\\
& = &
{\Gamma\left({1\over 4}\right)\over 4\sqrt{\pi}}
{1\over\beta^{3/4}\hbar}.
\label{ref1}
\eea
The kernels corresponding to rotations by $\pi/2$ and $\pi$ are
trivial since all integrals are done by delta functions leaving
\be \label{rot1}
Z(R_{\pi/2};\beta) = {1\over 2} \hspace{8ex}
Z(R_\pi;\beta)     = {1\over 4}.
\ee
We now go back and analyse in greater detail the first two integrals.
The first was studied by Tomsovic \cite{tom} but for completeness and 
consistency of notation, we review the calculation. 

Deep in one of the channels, $x\gg 1$ for example, 
the approximation using the Wigner transforms breaks down. This can be
understood as follows.
This approximation assumes that for short times one is free to
ignore the dynamics so that the calculation involves only the local
value of the Hamiltonian. Usually this is not problematic, however
one finds here that the channel effects violate this assumption. This
is because in one of the channels, $x\gg 1$ for example, we can treat the
problem adiabatically so that in the y
direction there is harmonic motion with a frequency $\omega_x=\sqrt{2}x$.
This frequency becomes arbitrarily large in the channel and there is no
time scale over which the dynamics can be ignored. We overcome this
problem by using an alternate representation of the heat kernels based
on the approximate separation of the problem as introduced in
Ref.~\cite{tom} and which we discuss below. This is a
complementary representation which is valid deep in the channels but fails
near the origin. To proceed, we assume that there is a domain of $x$ 
where both representations are valid. Let $Q$ be a value of
$x$ in this domain. The condition for the adiabatic representation to
be valid is that we be deep in one of the channels,
in terms of dimensionless quantities this is $\beta^{1/4}Q\gg 1$.
The condition for the Wigner function
representation to be valid is $\beta\hbar Q\ll 1$. These conditions
are compatible if $\beta^{3/4}\hbar\ll 1$. (If we determine all quantities
in units of energy $[E]$, then $Q$ has units of $[E^{1/4}]$, $\beta$ has 
units of $[E^{-1}]$ and $\hbar$ has units of $[E^{3/4}]$ so that all the
conditions mentioned above are in terms of dimensionless combinations.)
We will use the Wigner
representation in the square $|x|\leq Q$, $|y|\leq Q$ and the adiabatic
representation elsewhere. We then replace (\ref{Id}) by
\be \label{Id_1}
Z_0(I;\beta) = {1\over(2\pi\hbar)^2}{2\pi\over\beta}\int_{-Q}^Qdxdy
e^{-\beta x^2y^2}.
\ee
(We have introduced the subscript 0 to denote that this is the
contribution from the central region around the origin.)
This integral can be done by the change of variables
$u=\sqrt{\beta}xy$ and
$v=x$, so that the integrand is proportional to $\exp(-u^2)/v$. Doing
the $v$ integral first and using $\beta^{1/4}Q\gg 1$, one finds
\be \label{Id_1_res}
Z_0(I;\beta) =
\sqrt{{1\over\pi\beta^3\hbar^4}}\left(\log(2\sqrt{\beta}Q^2) +
{\gamma\over 2} \right),
\ee
where $\gamma=0.5772...$ is Euler's constant.
Similarly, for the reflection operator
$\sigma_y$, integration of (\ref{refy}) between the limits $-Q$ and $Q$
leads to 
\be \label{refy_1}
Z_0(\sigma_y;\beta) = \sqrt{{1\over 2\pi\beta\hbar^2}}Q.
\ee

To do the integrals in the channels we assume a local separation of the
Hamiltonian into a free particle in the $x$ direction and a 
harmonic oscillator in the $y$ direction, with a
frequency which depends parametrically on $x$
\be \label{separate}
h_x   =  {1 \over 2}p_y^2 + {\omega_x^2\over 2}y^2. 
\ee
Henceforth we will use small letters to denote objects related to the
local Hamiltonian $h_x$. It has eigenenergies $e_n=(n+1/2)\omega_x\hbar$
and eigenstates $|\phi_n\rangle$ which depend parametrically on $x$.
All the symmetry information to do with the channel calculation
is encoded in these local
eigenenergies and eigenstates. In particular, we are interested in the
subgroup of $C_{4v}$ which leaves $x$ invariant and so maps the local
eigenstates onto one another. This subgroup is just the parity group
with group elements $\{I,\sigma_y\}$.
This group has a trivial character table but we include it
for completeness as Table. 2.  For fixed $x$, we proceed in analogy to
(\ref{chk}) by defining heat kernels based on the local eigenvalues
and corresponding to these two group operations,
\bea
z_x(g,\beta) & = &
\mbox{tr}\left(\hat{U}^\dagger(g)e^{-\beta\hat{h}}\right) \nonumber \\
             & = & \sum_n \eta_n(g) e^{-\beta e_n} \label{lhk}
\eea
where $g$ is either the identity or the reflection element. The trace
operator ``tr'' denotes the local integral over the $y$ degree of freedom
and can be found by summing over the index $n$. It is clear that the
operator $\hat{U}^\dagger(g)$ is unity when $g=I$ and changes the sign
of the odd states when $g=\sigma_y$, so that $\eta_n(I)=1$ and
$\eta_n(\sigma_y)=(-1)^n$.

To evaluate the full trace, we note that the integrals in $p_y$ and $y$ have
already been done implicitly in (\ref{lhk}) so we only need to do the
$x$ and $p_x$ integrals. Since this is only one dimensional,
the prefactor of the integral has only one power of 
$2\pi\hbar$ and we conclude
\bea
Z_c(g;\beta) & = & f_g{1\over 2\pi\hbar}\int_{-\infty}^\infty dp_x 
e^{-\beta p_x^2/2}\int_Q^\infty dx z_x(g,\beta) \nonumber \\
             & = & f_g \sqrt{{1\over\pi\beta^3\hbar^4}}
\sum_n \eta_n(g){\xi^{2n+1}\over 2n+1}, \label{stuff}
\eea
where we have defined the factor $\xi=\exp(-\beta\hbar Q/\sqrt{2})$.
(We include a subscript $c$ to denote that this is the channel contribution.)
We have also introduced a factor $f_g$ which represents the number of channels
which map to themselves under the action of the group element $g$.
When working with the identity element, all the channels map onto themselves
and $f_I=4$; when working with the reflection operator the two
such channels along the $x$ axis map onto themselves and $f_{\sigma_y}=2$.
We now make use of the series identities
\bea
\sum_n{\xi^{2n+1}\over 2n+1} & = & {1\over 2}\log\left({1+\xi\over
1-\xi}\right)
\nonumber\\
\sum_n(-1)^n{\xi^{2n+1}\over 2n+1} & = & \arctan\xi 
\label{serid}
\eea
and the fact that $\beta\hbar Q\ll 1$ to conclude
\bea
Z_c(I;\beta) & = &
-\sqrt{{1\over\pi\beta^3\hbar^4}}\log\left({\hbar^2\beta^2Q^2\over
8}\right)\nonumber\\ 
Z_c(\sigma_y;\beta) & = & \sqrt{{\pi\over 4\beta^3\hbar^4}} - 
                          \sqrt{{1\over2\pi\beta\hbar^2}}Q.
\label{chann_both_res}
\eea
We add the results from inside the square
(\ref{Id_1_res}) and (\ref{refy_1}) to the channel results
(\ref{chann_both_res}) to get
\bea
Z(I;\beta) & = &
\sqrt{{1\over 4\pi\beta^3\hbar^4}}
\left(\log\left({1\over\beta^3\hbar^4}\right)+\gamma+8\log2\right) 
\nonumber\\
Z(\sigma_y;\beta) & = & \sqrt{{\pi \over 4\beta^3\hbar^4}}. \label{total}
\eea
Note that the $Q$ dependence has cancelled from both results 
leaving a finite answer. (This prescription actually overcounts some
regions of phase space but the errors so introduced are
exponentially small in $\beta^{1/4}Q$.)

We have now calculated the five class heat kernels which we need. All
that remains is to compute their inverse Laplace transforms. In fact, we
will not be interested in the densities $\rho(g;E)$ themselves but rather
in their integrals $N(g;E)$ which are given by
\be \label{ilt}
N(g;E) = {\cal L}^{-1}\left({Z(g;\beta)\over\beta}\right).
\ee
The inverse Laplace transforms are
\bea
N(I;E) &\  =\  & {2 \over 3\pi}y^2\left(4\log y + 4\gamma + 14\log2 -
8\right) \nonumber \\[1ex]
N(\sigma_y;E)  &\  =\  & {2 \over 3}y^2 \nonumber \\[1ex]
N(\sigma_1;E)  &\  =\  & {\Gamma^2({1\over 4}) \over \sqrt{18\pi^3}}y\nonumber\\[1ex]
N(R_{\pi/2};E) &\  =\  & {1 \over 2} \nonumber \\[1ex]
N(R_\pi;E)     &\  =\  & {1 \over 4}. \label{hwg}
\eea
We have defined the dimensionless scaled energy $y=E^{3/4}/\hbar$,
which is a semiclassically large quantity.  If we explicitly include
the mass $m$ in the kinetic energy of the Hamiltonian and a parameter
$\alpha$ in front of the potential energy then Eq.~(\ref{hwg}) still
applies but with  $y=(m^{1/2}E^{3/4})/(\alpha^{1/4}\hbar)$. We further
note that the inverse Laplace transforms imply that all the functions
are zero for negative energies.  The first of these relations is the
average integrated density of states summed over all irreps and was
already found by Tomsovic \cite{tom}. To construct the integrated
densities of states for each of the five irreps, we use
Eq.~(\ref{dodo}) with the symbols $Z$ replaced by the symbols $N$. It
should also be mentioned that these are just the leading order results
in an asymptotic semiclassical expansion. The terms in this series
will eventually diverge in a manner controlled by the shortest
periodic orbit \cite{berhowl}.

For typical two dimensional potentials with finite phase space
volumes, the term $N(I;E)$ scales as $1/\hbar^2$. The prefactor of
that term in (\ref{hwg}) has this scaling but there is a further
logarithmic dependence on $\hbar$ which causes it to grow somewhat
faster. This logarithmic factor arises from the fact that the integral
in (\ref{Id_1}) diverges logarithmically with $Q$. One must be careful
in discussing ``orders'' when expressions involve logarithms of large
quantities and for practical purposes, the non-logarithmic term
$4\gamma+14\log 2-8$ represents an essential correction, as
discussed in Ref.~\cite{tom}. Based on Eq.~(\ref{res}), we expect
terms involving reflection operators to be weaker by a relative power
of $\hbar$ and therefore to scale as $1/\hbar$. This is not true for
$N(\sigma_y;E)$ which is amplified by a
factor of $1/\hbar$ so that it is of the same order as the
non-logarithmic term in $N(I;E)$. The fact that it has been amplified
by a full power of $1/\hbar$ can be traced to the fact that the
integral (\ref{refy}) diverges linearly with $Q$. Therefore, rather
than being a relatively weak correction, this reflection operator is
almost leading order in its effect. In particular, the approximate
relation that the fraction of states in irrep $R$ is
approximately $d_R^2/|G|$ fails in general, since it comes from
considering just the identity operator. (However it is valid for the $E$ 
irrep which is independent of that reflection class.) A similar behaviour
is also apparent in the related problem of the hyperbola billiard 
\cite{hyp,dahl1}.
The other reflection
class function $N(\sigma_1;E)$ does scale as $1/\hbar$ as we expect for normal
reflection operations. The two rotation classes also
behave normally \cite{us}, being constants independent of $\hbar$.

\section{Numerical Comparison of Two Dimensional Results}

We have numerically diagonalised the quantum Hamiltonian and found the
first few hundred eigenvalues of the problem. We used appropriately
symmetrised bases involving harmonic oscillator wave functions in the
$x$ and $y$ directions to separately find the eigenvalues belonging to 
each irrep. All results are for bases of 200 oscillators in each direction.
To make the comparison more explicit, we convolved the numerically
obtained density of states by a Gaussian of width $w$,
\be \label{smooth}
\tilde{\rho}_R(E) = {1\over\sqrt{2\pi w^2}}\sum_n\exp{\left
(-{(E-E_n)^2\over 2w^2}\right)}.
\ee
The integrated density of states is then obtained by replacing the
sharp steps at the quantum eigenvalues by the corresponding error
functions. For large $w$, this convolution washes out all oscillations
leaving just the average behaviour.

In Fig.~\ref{irreps} we show the results for all five irreps with a
smoothing width $w=3$. The solid curves are the numerics and the
dashed curves are the analytical forms. The first thing which is
apparent is that there is a great distinction between the $A_1$ and
$B_1$ states compared to the $A_2$ and $B_2$ states, resulting from
the large contribution of $N(\sigma_y;E)$. Between each of these pairs
there is a much smaller splitting due to $N(\sigma_1;E)$. The
deviations between the solid and dashed curves are completely
numerical in origin and arise from the finite basis used in
determining the quantum eigenvalues. Due to the
channels, the eigenvalues converges very slowly with increased basis
size. It is interesting to note that the irreps which are odd with
respect to reflections through the channels are better converged.
Being odd, they are less sensitive to the effects of
the channels and are therefore less error prone. Nevertheless, their error is
still dominated by channel effects as we will demonstrate.
The other three irreps are not odd with respect to both
channels ($A_1$ and $B_1$ are even with respect to both channels
and the $E$ states can be chosen as even with
respect to one and odd with respect to the other.) All three of
them fail at approximately the same energy of $E\approx 18$. The
number of accurate eigenvalues is roughly 35 for $A_1$ and $B_1$
and 45 for $E$ (recall that $E$ is doubly degenerate so the number of
independent eigenvalues obtained is half the number of states plotted.)
This is rather dismal considering the 40,000 oscillator states used. 
The irreps $A_2$ and $B_2$ are accurate up to energies near $E\approx
60$ representing roughly 115 states each.

It is also interesting to numerically isolate the contributions from
the various classes and compare them to (\ref{hwg}) directly as done in
Ref.~\cite{us}. This is a simple exercise since the entries in the
character table are components of a unitary matrix which is readily
inverted. The result is
\be \label{invert}
\left(\begin{array}{c}
N(I;E)\\N(\sigma_y;E)\\N(\sigma_1;E)\\N(R_{\pi/2};E)\\N(R_\pi;E) 
\end{array}\right) = 
\left(\begin{array}{rrrrr}
 1& 1& 1& 1& 1\\
 1&-1& 1&-1& 0\\
 1& 1&-1&-1& 0\\
 1&-1&-1& 1& 0\\
 1& 1& 1& 1&-1
\end{array}\right)
\left(\begin{array}{c}
N_{A1}(E)\\N_{A2}(E)\\N_{B1}(E)\\N_{B2}(E)\\N_E(E)
\end{array}\right)
\ee
This can be written compactly as
\be \label{compact}
N(g;E) = \sum_R \eta_R(g) N_R(E)
\ee
where the factors $\eta_R(g)$ are defined in (\ref{invert}) and
can be thought of as the inverse of the group characters.
In Fig.~\ref{I_sx_r2} we plot $N(I;E)$, $N(\sigma_y;E)$ and
$N(R_\pi;E)$ from the theory and with the numerical eigenvalues
combined according to (\ref{invert}). 
As mentioned, the first is just the total number of states.
The third is shown in its own panel since its value is of
a very different scale than the other two.  They all fail around
$E\approx 18$ which is consistent with the previous 
figure. $N(R_\pi;E)$ depends on very fine cancellations
and is more sensitive to small errors so it is consistent that it
produces noticeable deviations at a slightly smaller energy than the other
two. Eq.~(\ref{hwg}) predicts a flat line for $N(R_\pi;E)$, the structure
at smaller $E$ comes from the convolution (\ref{smooth}) which is applied
to the analytical forms as well as to the numerical data.

The other two conjugacy classes behave very differently. We plot these
results in Fig.~\ref{s1_r1}. The upper panel shows $N(\sigma_1;E)$ and
the lower panel shows $N(R_{\pi/2};E)$. For the lower panel, we choose
two different smoothing widths, the relevance of which we discuss
below. For now, consider the comparison between the smooth solid curve
and the dashed curve in each case. The results are now accurate up to
energies of $E\approx 800$ or more than forty times the range
observed in the previous figure. This indicates that the numerics are,
in some sense, better than a quick study of Fig.~\ref{irreps} would
indicate. Although the various irreps are individually error prone
even at relatively modest energies, these errors are very correlated
so that appropriate combinations cause them to cancel. In fact, this is
apparent in Fig.~\ref{irreps} since the pairs $A_1$ and $B_1$ and also
$A_2$ and $B_2$ deviate from their expected behaviour
in very correlated manners. From (\ref{invert}) we see that both
$N(\sigma_1;E)$ and $N(R_{\pi/2};E)$ involve the differences
$N_{A1}(E)-N_{B1}(E)$ and $N_{A2}(E)-N_{B2}(E)$ and the systematic
effects cancel for these two classes. Since these functions agree with the
numerics up to $E\approx 800$, it is reasonable to associate all the problems
in the numerics with $N(I;E)$ and $N(\sigma_y;E)$, i.e. with the channels.
This is obviously true for the irreps $A_1$, $B_1$ and $E$, however it is
also true for the odd irres $A_2$ and $B_2$. Their staircase functions fail
at $E\approx 60$ which is better than the other irreps but still very
much smaller than the classes $N(\sigma_1;E)$ and $N(R_{\pi/2};E)$.

We now briefly discuss the oscillatory structure visible in the bottom
panel of Fig.~\ref{s1_r1}. This type of structure was also observed in
Ref.~\cite{us} where it was explained in terms of fractions of
periodic orbits \cite{robbins}. In this example, the structure arises 
from the square-like periodic orbit shown in Fig.~\ref{system}. After
completing, one quarter of a cycle, the trajectory is related to its
initial point by a rotation of angle $\pi/2$. This quarter-orbit then
contributes an oscillatory contribution to the function
$N(R_{\pi/2};E)$. This is a scaling system whose classical mechanics
is independent of energy, after appropriate scalings. In particular, 
the period of an orbit scales as $T\propto E^{-1/4}$ which
explains the growing wavelength with energy. Additionally, the smoothing
suppresses the oscillatory contribution by a factor proportional to 
$\exp{(-w^2 T^2/2)}$ which explains why the amplitude of oscillation
increases with energy. At the highest end of the energy range, one sees the
contributions of higher repetitions - for example three quarters of
the square orbit will also contribute to $N(R_{\pi/2};E)$. The
function $N(R_{\sigma_1};E)$ receives contributions from fractional
orbits which map to themselves under reflection through the diagonal.
Examples of this include the diagonal orbit after a half period and after a
full period. Such structure is visible at
the upper end of the energy range but is less apparent than in the
bottom panel because of the different vertical scale.

Similar structure exists for the other classes as well but is not
visible due to the short energy range available. $N(R_{\pi};E)$
receives contributions from one half the diagonal orbit and one half
the square orbit. $N(\sigma_y;E)$ receives a strong contribution from
the almost periodic family of orbits corresponding to the adiabatic
oscillation deep in the channels (actually, from the fractional
periodic family which has one half the period.) This is a non-standard
contribution due to the intermittency, such effects are discussed in
Refs.\cite{dahl1,dahl2,helium,hydrogen}. The function $N(I;E)$ receives
contributions from all the complete orbits but not from any fraction
of them. The periodic orbit theory of this system has been discussed
in detail in Ref.~\cite{dahl2} and the references therein, so we
forego a more detailed discussion.

\section{The three dimensional generalisation}

In this section we discuss the three dimensional potential
$V=x^2y^2+y^2z^2+z^2x^2$. This is the potential which actually appears
in the zero dimensional limit of the $SU(2)$ Yang-Mills equations.
The symmetry group is that of the octahedral group in which we allow
spatial inversions --- the extended octahedral group. In Fig.~\ref{3dpot} 
we show a three dimensional constant energy contour of the potential 
and also an octahedron whose vertices are aligned along the channel
directions. In total there are 48 group elements organised into 10
conjugacy classes. This group is the direct product of the
inversionless octahedral group and the inversion parity group. The
first of these is composed of 24 group elements organised into 5
classes \cite{lomont} and we start by enumerating these. First, there
is the identity $I$, which is in a class by itself. There is a class
of six elements involving rotations by $\pm\pi/2$ about any of the
three axes, such as $R_{x,\pi/2}$. Similarly, there is a class of
three elements involving rotations by $\pi$ about these axes, for
example $R_{x,\pi}$. There is a class of 8 elements involving rotation
by $\pm 2\pi/3$ about any of the face-face axes, such as $R_{a,2\pi/3}$.
Finally, there is a class of six elements involving rotations by $\pi$
about any of the the six edge-edge axes, such as $R_{1,\pi}$. We refer
to these classes as $C_1$ to $C_5$ respectively. This group has 5
irreps and the character table is the top left quarter of Table~3.

To construct the full group, we multiply representative members of 
each class by the the inversion operation $\Sigma=\sigma_x\sigma_y\sigma_z$.
The effect of this is to map the identity to the inversion element
$\Sigma$ and to map each rotation into either a single reflection or into a
rotation times a reflection. This induces five additional classes. The
element $\Sigma$ is in a class by itself. Composition of the second
class with $\Sigma$ gives a class of six elements which are rotations
by $\pm\pi/2$ through an axis times reflection through that axis, such
as $R_{x,\pi/2}\sigma_x$. Composition of the third class with
$\Sigma$ gives the reflection elements about the three planes, such as
$\sigma_x$. The fourth class becomes a product of a rotation about a
face-face axis times a reflection through the perpendicular plane,
such as $R_{a,2\pi/3}\sigma_a$. Finally, the fifth class becomes
reflections through planes defined by the edges and vertices. An
example is the plane defined by the point $1$ together with the vertices at
positive and minus $z$. We call reflections through this plane $\sigma_1$.
We denote these five additional classes $C_1'$ to $C_5'$ respectively.
The addition of these classes doubles the number of irreps and the
full character table is shown in Table~3.

We proceed by analogy with the two dimensional problem. There we found
that to analyse the contribution of a single channel, it was necessary
to consider the subgroup which mapped that channel onto itself --- in
that case it was the parity group. We do the same here. The eight group
elements which map the channel $x\gg 1$ (for example) onto itself are
$I$, $\sigma_{y,z}$, $\sigma_{2,3}$, $R_{x,\pm\pi/2}$ and $R_{x,\pi}$
and these belong to classes $C_1$, $C_3'$, $C_5'$, $C_2$ and $C_3$
respectively. ($\sigma_{2,3}$ are defined in analogy to $\sigma_1$;
they are reflections through the two planes defined by the vertices
at plus and minus $x$ and the midpoints of the two edges connecting the
$z$ vertex to the positive and negative $y$ vertices.) We can expect the
integrals associated with these elements to be problematic and to
possibly require the adiabatic matching used in section III. 
Together these eight elements comprise the subgroup $C_{4v}$ which is,
of course, the group we studied in the two dimensional problem.
This will prove useful in the subsequent analysis. 

We start by studying the five classes which do not require an
adiabatic analysis. The class $C_1'$ involves three orthogonal
reflections while the classes $C_2'$ and $C_4'$ involve rotations and
perpendicular reflections. Their Wigner transforms are given by
(\ref{3dres}) and are trivial to integrate since they involve delta
functions of all the quantities. Their constributions are $1/8$, $1/4$
and $1/6$ respectively. The class $C_4$ involves rotations through the
face axes. For rotation by $2\pi/3$ through the point $a$, we define a
change of variables
\be
\xi   = {1\over\sqrt{3}}( x - y + z) \hspace{1cm} 
\eta  = {1\over\sqrt{6}}(2x + y - z) \hspace{1cm}
\zeta = {1\over\sqrt{2}}(y+z), \label{cov_4}
\ee
so that the potential along the $\xi$ axis is $V=\xi^4/3$. We then use
the third equation of (\ref{res}) with this choice of variables to find
\be \label{c_4}
Z(C_4;\beta) = { \Gamma({1\over 4}) \over \sqrt{24\sqrt{3}\pi} }
{1\over\beta^{3/4}\hbar }.
\ee
For rotation by $\pi$ through the point $1$, we define a change of
variables
\be
\xi   = {1\over\sqrt{2}}(x + y) \hspace{1cm}
\eta  = {1\over\sqrt{2}}(x - y) \hspace{1cm}
\zeta = z                       \label{cov_5}
\ee
so that the potential along the $\xi$ axis is $V=\xi^4/4$. We then
find
\be \label{c_5}
Z(C_5;\beta) = {\Gamma({1\over 4})\over 8\sqrt{\pi}}
{1\over\beta^{3/4}\hbar}.
\ee

We now consider the more interesting classes which map at least one
channel onto itself. We earlier suggested that the integrals
corresponding to them might be problematic. In fact, this is true for
all of them except the identity whose integral converges without
such an analysis. Therefore, we do it first,
\bea
Z(I;\beta) & = & {1\over(2\pi\hbar)^3} \int dxdydzdp_xdp_ydp_ze^{-\beta H}
\nonumber\\
           & = & {\Gamma^3({1\over 4})\over \sqrt{32\pi^3}}
{1\over\beta^{9/4}\hbar^3}.
\label{c_1}
\eea
%{1\over(2\pi\hbar)^3}\left({2\pi\over\beta}\right)^{3/2}
%\int_{-\infty}^{\infty}dxdy\sqrt{{\pi\over\beta(x^2+y^2)}} 
%e^{-\beta x^2y^2}. 
(The $p$ integrals are done trivially and the spatial integrals can be done
by a change to cylindrical coordinates.)
The convergence of this integral is due to the fact that deep in one
of the channels, the energetically accessible area pinches off as $1/x^2$,
which is integrable. The analogous integral in two dimensions pinches
off as $1/x$ and is not integrable. The remaining four classes follow
from using (\ref{trwt}), (\ref{res}) and (\ref{chk}) inside a cube 
$|x|\leq Q$, $|y|\leq Q$ and $|z|\leq Q$. 

For reflection in $z$, which is a member of the $C_3'$ class,
we use Eq.~(\ref{res}) and so arrive at the following integral,
\be \label{c_3'_0a}
Z_0(\sigma_z;\beta) = {1\over 2} {1\over (2\pi\hbar)^2} 
\int_{-\infty}^\infty dp_xdp_y e^{-\beta(p_x^2+p_y^2)/2}
\int_{-Q}^Q dxdy e^{-\beta x^2y^2}.
\ee
Other than the factor of one half, this is the same integral
we evaluated to get the total density of states in the two dimensional
problem. The result is given by (\ref{Id_1_res}) and so we conclude
\be \label{c_3'_0}
Z_0(\sigma_z;\beta) = \sqrt{{1\over\pi\beta^3\hbar^4}}
\left(\log Q + \log\beta^{1/4} + {\gamma\over 4} + {1\over 2}\log 2\right).
\ee
Reflection in $\sigma_3$, which is a member of the $C_5'$ class,
requires a more complicated calculation. We define a change of
coordinates so that $\eta = (z+y)/\sqrt{2}$ and $\zeta =
(z-y)/\sqrt{2}$ and then use Eq.~(\ref{res}) with the delta functions
acting on $\zeta$ and $p_\zeta$ so that the integral to be evaluated is
\be \label{c_5'_0a}
Z_0(\sigma_3;\beta) = {1\over \pi\beta\hbar^2}
\int_0^Qdx\int_0^{\sqrt{2}Q}d\eta
e^{-\beta(x^2\eta^2+\eta^4/4)}.
\ee
We have done the trivial momentum integrals and have noted that by its
definition, $\eta$ has a different integration range than $x$. This
integral can be done in a manner analogous to (\ref{Id_1}), we define
integration variables $u=x\eta$ and $v=\eta$. Doing the $v$
integration first and using $\beta^{1/4}Q\gg 1$ one
arrives at
\be \label{c_5'_0}
Z_0(\sigma_3;\beta) = \sqrt{{1\over 4\pi\beta^3\hbar^4}}
\left(\log Q + \log\beta^{1/4} +{\gamma\over 4} + {3\over 2}\log 2\right).
\ee
 
Rotation by $\pi/2$ about the $x$ axis is a member of the $C_2$ class
and implies delta functions in the other two variables so that the
integral to be done is
\bea
Z_0(R_{\pi/2};\beta) & = & {1\over 4\pi\hbar}
\int_{-\infty}^\infty dp_x e^{-\beta p_x^2/2}\int_{-Q}^Q dx \nonumber \\
		     & = & \sqrt{{1\over 2\pi\beta\hbar^2}}Q.
\label{c_2_0}
\eea
Rotation by $\pi$ about the $x$ axis, which is a member of the $C_3$
class, involves an integral which is identical except for a factor
of two from the $\sin^2(\theta/2)$ factor in (\ref{res}). Therefore
\be \label{c_3_0}
Z_0(R_\pi;\beta) = \sqrt{{1\over 8\pi\beta\hbar^2}}Q.
\ee

\section {Channel Calculations in Three Dimensions}

In this section we evaluate the contribution of the channels in three
dimensions. As discussed before, this is is only necessary for some of
the group elements. In analogy with
(\ref{separate}) we define a local two-dimensional Hamiltonian as
\be \label{separate_3d}
h_x   =  {1 \over 2}(p_y^2 + p_z^2) + {\omega_x^2\over 2}(y^2+z^2) + y^2z^2,
\ee
where again $\omega_x=\sqrt{2}x$ and $x$ is assumed large. 
Deep in the channel, the
final term can be thought of as a small perturbation which has
virtually no effect on the eigenenergies. If that term were completely
absent, the local Hamiltonian would have an $SU(2)$ symmetry
corresponding to a two-dimensional harmonic oscillator. The
eigenvalues of the Hamiltonian would then be $e_n=(n+1)\hbar\omega_x$,
each with a degeneracy of $(n+1)$. The degenerate states can be
labelled by the rotational quantum number $m$ which runs from $-n$ to
$n$ in even increments. The perturbation $y^2z^2$ will not affect the
energies in a significant manner but will act to break up the
degenerate collections of states into specific irreps of $C_{4v}$ as
follows. All states with odd $m$ correspond to the $E$ irrep. 
The $m=0$ states are all $A_1$. For $m$ non-zero and divisible by 4, the
states are either $A_1$ or $B_2$ (corresponding to $\cos(m\theta)$ and
$\sin(m\theta)$ respectively). Otherwise, if $m$ is even but not
divisible by $4$, the states are either $A_2$ or $B_1$ (corresponding
to $\sin(m\theta)$ and $\cos(m\theta)$ respectively.) We then define
local heat kernels corresponding to the five irreps by adding
the contributions of all values of $n$ with the appropriate degeneracy
factor for each irrep so that
\bea
z_{A_1}(\beta)  & =  & \sum_{n=\mbox{even}} \left[{n+4 \over 4}\right] 
e^{-\beta\hbar\omega_x(n+1)} \nonumber\\[1ex]
z_{B_2}(\beta)  & =  & \sum_{n=\mbox{even}} \left[{n\over 4}\right] 
e^{-\beta\hbar\omega_x(n+1)} \nonumber\\[1ex]
z_{B_1}(\beta) = z_{A_2}(\beta)  & =  & \sum_{n=\mbox{even}}
\left[{n+2\over 4}\right] e^{-\beta\hbar\omega_x(n+1)} \nonumber\\[1ex]
z_{E}(\beta)    & =  & \sum_{n=\mbox{odd}}  
(n+1) e^{-\beta\hbar\omega_x(n+1)}, 
\label{lhk_3d}
\eea
where $[x]$ is the largest integer less than or equal to $x$. We will
refer to these relations collectively as
\be \label{compacter}
z_R(\beta) = \sum_{n=0}^\infty c_R(n)e^{-\beta\hbar\omega_x(n+1)},
\ee
where $c_R(n)$ are the degeneracy factors defined in (\ref{lhk_3d}).

To evaluate the traces, we integrate over the remaining $x$ dependence
\bea
Z_R(\beta) & = & {1\over 2\pi\hbar}\sum_n c_R(n)
\int_{-\infty}^\infty dp_x e^{-\beta p_x^2/2}
\int_Q^\infty dx e^{-\beta\hbar\sqrt{2}(n+1)x}\nonumber\\
           & = & \sqrt{{1\over 4\pi\beta^3\hbar^4}}\sum_n c_R(n)
{\xi^{n+1} \over n+1},
\label{Z_R_gen}
\eea
where we have defined $\xi=\exp{(-\sqrt{2}\beta\hbar Q)}
\approx 1-\sqrt{2}\beta\hbar Q$. (Note that this is different by a
factor of two from the analogous variable in two dimensions.)
All of this discussion is in terms of the local irreps; what we really
want, however, are the local class heat kernels. These we can get by
appropriate combinations of the irreps as in (\ref{compact}) to arrive
at the class sums
\be \label{classsum}
S(g;\beta) = \sum_n c(g,n){\xi^{n+1}\over n+1}.
\ee
For the moment we omit the prefactor of (\ref{Z_R_gen}), this will be
reintroduced later. The degeneracy factor $c(g,n)$ corresponding to a
group element $g$ is found by adding together the degeneracy factors
$c_R(n)$ with the appropriate weightings as given by (\ref{invert}), i.e.
\be \label{dgn}
c(g,n) = \sum_R \eta_R(g)c_R(n).
\ee

We start with the identity element. Earlier it was argued that we do
not need a channel calculation since the central integration
converges. However, it is of interest to see how this is also apparent
in the channel calculation. Comparing (\ref{invert}) and (\ref{lhk_3d})
it is apparent that $c(I,n)=n+1$ so that
\bea
S(I;\beta) = \sum_{n=0}^\infty \xi^{n+1} & = & {\xi\over 1-\xi}\nonumber\\
           & = & {1\over \sqrt{2}\beta\hbar Q}.
\label{sum_Id_chann}
\eea
We now need to reinsert the prefactor of (\ref{Z_R_gen}) and must also
include an integral
factor representing the number of channels left invariant by the
corresponding element $f_g$, as in two dimensions. We
trivially have $f_I=6$ so that the channel result for the identity
element is
\be \label{Id_chann}
Z_c(I;\beta) = {3\over\sqrt{2\pi}} {1\over \beta^{5/2}\hbar^3Q}.
\ee
We now compare this result to (\ref{c_1}), the present contribution is
very much smaller if $\beta^{1/4}Q\gg 1$ which is precisely the
limit we are considering. Therefore, we again observe that no channel
calculation is necessary for the identity element.

We next consider the reflection element $\sigma_z$. It is in the same class
as $\sigma_y$ so comparing (\ref{invert}) and (\ref{lhk_3d}) we conclude
$c(\sigma_z,n)=1$ when $n$ is even and $0$ when $n$ is odd. The sum which 
must be done is
\bea
S(\sigma_z;\beta) = \sum_{n=\mbox{even}}{\xi^{n+1}\over n+1}
                  & = & \sum_{m=0}^\infty {\xi^{2m+1}\over 2m+1}\nonumber\\
                  & = & {1\over 2} \log\left({\sqrt{2}\over \beta\hbar Q}\right),
\label{sum_refz_chann}
\eea
where we have used (\ref{serid}) and the approximation immediately
below (\ref{Z_R_gen}). Note that $f_{\sigma_z}=4$
since $\sigma_z$ leaves four channels invariant, so that
\be \label{C_3'_chann}
Z_c(\sigma_z;\beta) = \sqrt{{1\over \pi\beta^3\hbar^4}}
\left(\log{\sqrt{2}\over\beta\hbar} - \log Q\right).
\ee
Recalling now the corresponding result for the central region 
(\ref{c_3'_0}), we conclude that for the class $C_3'$,
\be \label{C_3'_finally}
Z(C_3';\beta) = \sqrt{{1\over 16\pi\beta^3\hbar^4}}
\left(\log\left({1\over\beta^3\hbar^4}\right) + \gamma + 4\log 2\right).
\ee
This is independent of $Q$ as we expect.

The equality of $z_{B_1}$ and $z_{A_2}$ in (\ref{c_3'_0}) implies that
$S(\sigma_3;\beta)=S(\sigma_z;\beta)$ (since they both equal
$z_{A_1}-z_{B_2}$ from (\ref{invert}).) The only difference is the
subsequent calculation is that $f_{\sigma_3}=2$ so that the channel
calculation is one half of that for $\sigma_z$ (\ref{C_3'_chann}). We
combine this result with the result from the central region (\ref{c_5'_0})
to determine 
\be \label{C_5'_finally}
Z(C_5';\beta) = \sqrt{{1\over 64\pi\beta^3\hbar^4}}
\left(\log\left({1\over\beta^3\hbar^4}\right) + \gamma +8\log 2\right).
\ee

For rotations by $\pi/2$ about the $x$ axis we note that 
$c(R_{\pi/2},n)=(-1)^{n/2}$ for $n$ even and is $0$ for $n$ odd
so that
\bea
S(R_{\pi/2};\beta) = \sum_{n=\mbox{even}}(-1)^{n/2}{\xi^{n+1}\over n+1}
             & = &  \sum_{m=0}^\infty (-1)^m{\xi^{2m+1}\over 2m+1}\nonumber\\
             & = & \arctan\xi,
\label{sum_rot1_chann}
\eea
where we have again used (\ref{serid}). We now note that
$\arctan\xi\approx \pi/4 -\beta\hbar Q/\sqrt{2}$ and also that only
two channels are left invariant implying $f_{R_{\pi/2}}=2$ so that
\be \label{C_2_chann}
Z_c(R_{\pi/2};\beta) = \sqrt{\pi\over 16\beta^3\hbar^4} 
- \sqrt{{1\over 2\pi\beta\hbar^2}}Q.
\ee
We now combine this with the calculation from the central region
(\ref{c_2_0}) to arrive at
\be \label{C_2_finally}
Z(C_2;\beta) = \sqrt{{\pi\over 16\beta^3\hbar^4}}.
\ee

The final class to be analysed is $C_3$ of which rotations about the
$x$ axis by $\pi$ is a representative member. We now have
$c(R_\pi,n)=(-1)^n$ so that the relevant sum is
\bea
S(R_\pi;\beta) & = & \sum_{n=0}^\infty (-1)^n\xi^{n+1} = {\xi\over
1+\xi} \nonumber \\
& = & {1\over 2} \left( 1-{\beta\hbar \over \sqrt{2}}Q\right).
\label{sum_rot2_chann}
\eea
As for the previous rotation class, we have $f_{R_\pi}=2$ so that
the result of the channel calculation is
\be \label{C_3_chann}
Z_c(R_\pi;\beta) = \sqrt{{1\over 4\pi\beta^3\hbar^4}} - 
                   \sqrt{{1\over 8\pi\beta\hbar^2}}Q.
\ee
Combining this with the calculation from the central region
(\ref{c_3_0}) we conclude
\be \label{C_3_finally}
Z(C_3;\beta) = \sqrt{{1\over 4\pi\beta^3\hbar^4}}.
\ee

The final analysis we will do is to find the inverse Laplace transform 
of the various relations and thereby express them in the energy domain.
The ten results as a function of $\beta$ are scattered over the
previous two sections. As in the two dimensions, we go directly
to the integrated densities of states by use of (\ref{ilt}). The result is
\bea
N(C_1;E)  & \ =\  & {16\Gamma^2({1\over 4})\over45\sqrt{2\pi^3}}y^3      \nonumber\\[1ex]
N(C_2;E)  & \ =\  & {1\over 3}    y^2                                   \nonumber\\[1ex]
N(C_3;E)  & \ =\  & {2\over 3\pi} y^2                                   \nonumber\\[1ex]
N(C_4;E)  & \ =\  & {\Gamma^2({1\over 4})\over\sqrt{27\sqrt{3}\pi^3}} y \nonumber\\[1ex]
N(C_5;E)  & \ =\  & {\Gamma^2({1\over 4})\over 6\sqrt{2\pi^3}}        y \nonumber\\[1ex]
N(C_1';E) & \ =\  & {1\over 8}                                          \nonumber\\[1ex]
N(C_2';E) & \ =\  & {1\over 4}                                          \nonumber\\[1ex]
N(C_3';E) & \ =\  & {1\over 3\pi} y^2 (4\log y+4\gamma+10\log 2-8)      \nonumber\\[1ex]
N(C_4';E) & \ =\  & {1\over 6}                                          \nonumber\\[1ex]
N(C_5';E) & \ =\  & {1\over 6\pi} y^2 (4\log y+4\gamma+14\log 2-8),
\label{heretheyare}
\eea
where again we use the semiclassically large quantity $y=E^{3/4}/\hbar$.
For comparison, we remark that for generic potentials, use of
(\ref{res}), would imply that the first term scales as $y^3$, the
following four as $y$, the set $\{C_1',C_2',C_4'\}$ as $y^0$, and the
set $\{C_3',C_5'\}$ as $y^2$.

The leading order behaviour, as given by the first expression, scales
generically with $\hbar$. There are no other terms
which are competitive with it so the relation that
the fraction of states in irrep $R$ is approximately $d_R^2/|G|$ is
valid. As discussed in the text the reflection classes $C_3'$ and
$C_5'$ are amplified somewhat, having an additional logarithmic
dependence on $\hbar$ in addition to the $1/\hbar^2$ prefactor. This
is in analogy to the total density of states of the two dimensional
problem. In fact, the class $C_5'$ is, within a factor of four, the same
as the total density of states in two dimensions.
Two of the rotation classes are amplified by $1/\hbar$ so that they 
scale as $1/\hbar^2$. This makes them competitive with the reflection
classes (since, as argued in the two dimensional problem, the logarithmic
term is a rather weak amplification). This is analogous to the
behaviour of one of the reflection operators in the two dimensional
case. 

In Fig.~\ref{3dresults} we show the integrated densities of states 
found from using the results of (\ref{heretheyare}) combined according
to the characters of Table~III.
(It should be remarked that this is not entirely consistent since the
leading order terms have semiclassical corrections which are almost
certainly of the same order or larger than the smallest terms we are
considering. However, the point of this paper is not a systematic
semiclassical expansion but rather a study of the symmetry effects.) 
The structure now looks more typical; irreps of the same
dimensionality have roughly similar numbers of states with slight
differences arising from the contributions of the other group
elements. In particular, the largest four curves are the four three
dimensional irreps and the differences among them arise from the terms
of order $y^2\log y$ and $y^2$; the largest of these curves  belongs
to $\Gamma_5'$. The middle two curves belong to the two dimensional
irreps and the smallest four curves belong to the one dimensional
irreps. The largest of these is the trivial irrep $\Gamma_1$; this is
reasonable since it receives positive contributions from all the classes.

In Fig.~\ref{more3dresults} we show the same data but on a smaller
energy scale. At the right edge of the figure ($E=35$), the curves are
ordered the same as in Fig.~\ref{3dresults} (i.e. their asymptotic ordering).
However, it is clear that there is a lot of crossing of these curves
at lower energies. This is because for moderate energies the
contribution corresponding to identity in (\ref{heretheyare}) does not
dominate the others. Additionally, in calculating the functions for
each irrep via (\ref{dodo}) (with the symbol $Z$ replaced by $N$), we
must sum over all the group elements and so the contribution of any
given class is amplified by the number of elements in that class. The
identity class only has one element but the classes which contribute
to next order, $\{C_2,C_3,C_3',C_5'\}$, have six, three, three and six
elements respectively. As mentioned, it is difficult to calculate many
accurate eigenvalues when a potential has channels and this is especially
true in three dimensions. Therefore, the non-asymptotic behaviour in
Fig.~\ref{more3dresults} is relevant to any numerical study since
the results will probably all be in that energy domain.

\section*{Conclusion}

We have shown that the symmetry reduction of the
Thomas-Fermi density of states discussed in Ref.~\cite{us} is easily 
generalised to more perverse systems where the Wigner representation fails.
In two dimensions, the symmetry decomposition introduces
essentially leading  order contributions to the densities of states of
the one dimensional irreps. The results were verified numerically and
seen to work well. However, the problem studied is numerically very
difficult and only a  handful of states of each irrep are reliably
calculated. Nevertheless, certain combinations of the densities of
states are found to be accurate to very high energies even though the
density of states of each individual irrep is not. This effect is
noticeable only by studying the class functions derived here and
would not otherwise have been apparent, thus underlining the
importance of symmetry decompositions.

In three dimensions, we find that the symmetry decomposition
does not introduce terms which are essentially leading order. However,
there are still interesting effects; two of the reflection classes
have a logarithmic dependence on $\hbar$ beyond what one might have
expected and two of the rotation classes have an additional power of
$1/\hbar$ thus making them of essentially the same order as the
reflection elements. Furthermore, we observed that even in this case
one must consider rather high energies before the ordering of the
functions $N_R(E)$ achieves its final form. This is in spite of the
fact that the leading behaviour is not affected by the decomposition.
Rather it arises from the fact that the classes which contribute at
next to leading order have several group elements and their
contributions are correspondingly amplified. This is an effect which
we can expect to become even more important in higher dimensions if we
consider potentials of the form  $V(\{x_i\}) = \sum_i\sum_{j>i}x_i^2x_j^2$.
In higher dimensions, more and more of the terms will behave with the normal
$\hbar$ dependence. The only terms with anamalous dependences are
those for which one would initially expect a dependence of $1/\hbar^2$
or $1/\hbar$. If the corresponding group element leaves at least one
channel invariant, they will be amplified by factors of
$\log(1/\hbar)$ and $1/\hbar$ respectively.

\begin{acknowledgements}
The author would like to thank Stephen Creagh and Bent Lauritzen for
useful discussions and the National Sciences and Engineering Research
Council of Canada for support.
\end{acknowledgements}

\begin{figure}
\caption{The configuration space of the $x^2y^2$ potential. The light
solid lines are constant energy contours at $E=0.1,1,3,5,\cdots$ and
the four dashed lines indicate the axes through which the system has a
reflection symmetry. The two heavy curves show
the two shortest periodic orbits in the system calculated at $E=1$.}
\label{system}
\end{figure}

\begin{figure}
\caption{The solid curves indicate the smoothed density of states for
each of the five irreps as found numerically. The dashed curves are
the corresponding analytical forms derived in this paper.}
\label{irreps}
\end{figure}

\begin{figure}
\caption{Top: The upper pair of curves indicates the function $N(I;E)$
which is the theory for the total density of states. The lower pair
shows $N(\sigma_y;E)$. In each case the solid curve comes from the
numerics and the dashed curve is the analytical form. Bottom: the same
for $N(R_\pi;E)$.}
\label{I_sx_r2}
\end{figure}

\begin{figure}
\caption{The same as the previous figure except that the upper panel
indicates $N(\sigma_1;E)$ and the lower panel is $N(R_{\pi/2};E)$. The
solid oscillating curve has with a smoothing width of 3. The other
solid curve and analytical dashed curve have smoothing widths of 30.}
\label{s1_r1}
\end{figure}

\begin{figure}
\caption{
Left: An equal energy contour of the three dimensional potential 
$V=x^2y^2+y^2z^2+z^2x^2$ showing the six channels along the three axes.
Right: An octahedron with the relevant points labelled for the
description of the group elements.}
\label{3dpot}
\end{figure}

\begin{figure}
\caption{
The average densities of states for each of the ten irreps of the 
$V=x^2y^2+y^2z^2+z^2x^2$ potential. From greatest to smallest the curves
describe the irreps $\Gamma_5'$, $\Gamma_4$, $\Gamma_4'$, $\Gamma_5$,
$\Gamma_3$, $\Gamma_3'$, $\Gamma_1$, $\Gamma_2'$, $\Gamma_2$ and $\Gamma_1'$.}
\label{3dresults}
\end{figure}

\begin{figure}
\caption{
The same as Fig.~6 but on a smaller energy scale to show the curves
crossing at small energies. At $E=35$, the order of the curves is the
same as that described in the previous figure caption.}
\label{more3dresults}
\end{figure}

\begin{table} \label{tab1}
\begin{center}
\begin{tabular}{rrrrrrrrrrr}
  & $I$ & $\sigma_{x,y}$ & $\sigma_{1,2}$ & $R_{\pm\pi/2}$ & $R_\pi$\\
\hline
$A_1$ \vline & $1$ &  $1$  &   $1$  &   $1$   &   $1$  \\
$A_2$ \vline & $1$ & $-1$  &   $1$  &  $-1$   &   $1$  \\
$B_1$ \vline & $1$ &  $1$  &  $-1$  &  $-1$   &   $1$  \\
$B_2$ \vline & $1$ & $-1$  &  $-1$  &   $1$   &   $1$  \\
$E$   \vline & $2$ &  $0$  &   $0$  &   $0$   &  $-2$  \\
\end{tabular}
\caption{Character table of the group $C_{4v}$.}
\end{center}
\end{table}

\begin{table} \label{tab2}
%\begin{center}
\begin{tabular}{rrrrrrrrrrr}
  & $I$ & $\sigma_y$ \\
\hline
$E$ \vline & $1$ &  $1$ \\
$O$ \vline & $1$ & $-1$ \\
\end{tabular}
\caption{Character table of the parity group.}
%\end{center}
\end{table}

\begin{table} \label{tab3}
\begin{center}
\begin{tabular}{rrrrrrrrrrr}
  & $C_1$    & $C_2$  & $C_3$  & $C_4$  & $C_5$ 
  & $C'_1$   & $C'_2$ & $C'_3$ & $C'_4$ & $C'_5$ \\
 & (1) & (6) & (3) & (8) & (6) & (1) & (6) & (3) & (8) & (6)\\
\hline
$\Gamma_1$ \vline & $1$ &  $1$  &   $1$  &   $1$   &   $1$   & $1$ &  $1$  &   $1$  &   $1$   &   $1$\\
$\Gamma_2$ \vline & $1$ & $-1$  &   $1$  &   $1$   &  $-1$  & $1$ & $-1$  &   $1$  &   $1$   &  $-1$  \\
$\Gamma_3$ \vline & $2$ &  $0$  &   $2$  &  $-1$   &   $0$  & $2$ &  $0$  &   $2$  &  $-1$   &   $0$  \\
$\Gamma_4$ \vline & $3$ & $-1$  &  $-1$  &   $0$   &   $1$  & $3$ & $-1$  &  $-1$  &   $0$   &   $1$  \\
$\Gamma_5$ \vline & $3$ &  $1$  &  $-1$  &   $0$   &  $-1$  & $3$ &  $1$  &  $-1$  &   $0$   &  $-1$  \\
$\Gamma_1'$ \vline & $1$ &  $1$  &   $1$  &   $1$   &   $1$  & $-1$ &  $-1$  &   $-1$  &   $-1$   &   $-1$\\
$\Gamma_2'$ \vline & $1$ & $-1$  &   $1$  &   $1$   &  $-1$  & $-1$ & $1$  &   $-1$  &   $-1$   &  $1$  \\
$\Gamma_3'$ \vline & $2$ &  $0$  &   $2$  &  $-1$   &   $0$  & $-2$ &  $0$  &   $-2$  &  $1$   &   $0$  \\
$\Gamma_4'$ \vline & $3$ & $-1$  &  $-1$  &   $0$   &   $1$  & $-3$ & $1$  &  $1$  &   $0$   &   $-1$  \\
$\Gamma_5'$ \vline & $3$ & $1$   &  $-1$  &   $0$   &  $-1$  & $-3$ & $-1$ & $1$  &   $0$   &  $1$  \\
\end{tabular}
\caption{Character table of the extended octahedral group. The number in 
brackets at the top of each column indicates the number of group elements
which belong to that class. Representative members of the various classes 
are described in the text.}
\end{center}
\end{table}

\begin{references}

\bibitem[*]{cnrsbullshit} Unit\'{e} de recherche des Universit\'{e}s
de Paris XI et Paris VI associ\'{e}e au CNRS.

\bibitem{fft} B. M\"{u}ller and A. Trayanov, Phys. Rev. Lett. {\bf 68}, 3387
(1992); C. Gong, Phys. Rev. {\bf D49}, 2842 (1994); T. S. Bir\'{o} et. al.,
Int. J. Mod. Phys. {\bf C5}, 13 (1994); H. B. Nielsen, H. H. Rugh and S. E.
Rugh, ``Chaos and Scaling in Classical Non-Abelian Gauge
Fields'',chao-dyn/9605013; B. M\"{u}ller, ``Study of Chaos and Scaling in
Classical SU(2) Gauge Theory'', chao-dyn/9607001.

\bibitem{zerod} S. G. Matinyan, G. K. Savvidy and N. G.
Ter-Arutyunyan-Savvidy, Sov. Phys. JEPT {\bf 53}. 421 (1981); JETP Lett.
{\bf 34}, 590 (1982); B. V. Chirikov and D. L. Shepelyanskii, JETP Lett. 
{\bf 34}, 163 (1981); G. K. Savvidy, Nucl. Phys. {\bf B 246}, 302 (1984); G.
Berman, E. Bulgakov, D. Holm and Y. Kluger, Phys. Lett. {\bf A 194}, 251
(1994).

\bibitem{dahlruss} P. Dahlqvist and G. Russberg, Phys. Rev. Lett. {\bf
65}, 2837 (1990).

\bibitem{dahl1}  P. Dahlqvist, J. Phys. A: Math. Gen. {\bf 25}, 6265 (1992).

\bibitem{dahl2}  P. Dahlqvist, J. Phys. A: Math. Gen {\bf 27}, 763 (1994). 

\bibitem{helium} G. Tanner and D. Wintgen, Phys. Rev. Lett. {\bf 75}, 2928
(1995).

\bibitem{hydrogen} G. Tanner, K. T. Hansen and J. Main, to appear
Nonlinearity (1996).

\bibitem{eckwin} B. Eckhardt and D. Wintgen, J. Phys. B: At. Mol. Opt. Phys
{\bf 23}, 355 (1990).

\bibitem{simon}    B. Simon, Ann. Phys. {\bf 146}, 209 (1983).

\bibitem{tom}      S. Tomsovic, J. Phys. A: Math. Gen. {\bf 24}, L733 (1991).

\bibitem{pavloff}  N. Pavloff, J. Phys. A: Math. Gen {\bf 27}, 4317 (1994).

\bibitem{us}       B. Lauritzen and N. D. Whelan, Ann. Phys, 
{\bf 244}, 112 (1995).

\bibitem{btu}      O. Bohigas, S. Tomsovic and D. Ullmo, Phys. Rep.
{\bf 223}, 43 (1993).

\bibitem{sw}       H. M. Sommermann and H. A. Weidenm\"{u}ller,
Europhys. Lett. {\bf 23}, 79 (1993).

\bibitem{gutz}     M. C. Gutzwiller, {\it Chaos in Classical and Quantum 
                   Mechanics} (Springer Verlag, New York, 1990).
\bibitem{hamermesh} M. Hammermesh, ``Group Theory and its Applications'',
(Addison-Wesley, London, 1962).

\bibitem{jbb}      see for example, B. K. Jennings, R. K. Bhaduri and
M. Brack, Nucl. Phys. A {\bf 253}, 28 (1975) and references therein.

\bibitem{berhowl}  M.V. Berry and C. J. Howls, 
Proc. R. Soc. Lond. A {\bf 447}, 527 (1994).

\bibitem{hyp}      F. Steiner and P. Trillenberg, J. Math. Phys. {\bf
31}, 1670 (1990).

\bibitem{robbins}  J. Robbins, Phys. Rev. A {\bf 40}, 2128 (1989).

\bibitem{lomont}   see for example, J. S. Lomont, ``Applications of
Finite Groups'', (Academic Press, New York, 1959)

\end{references}
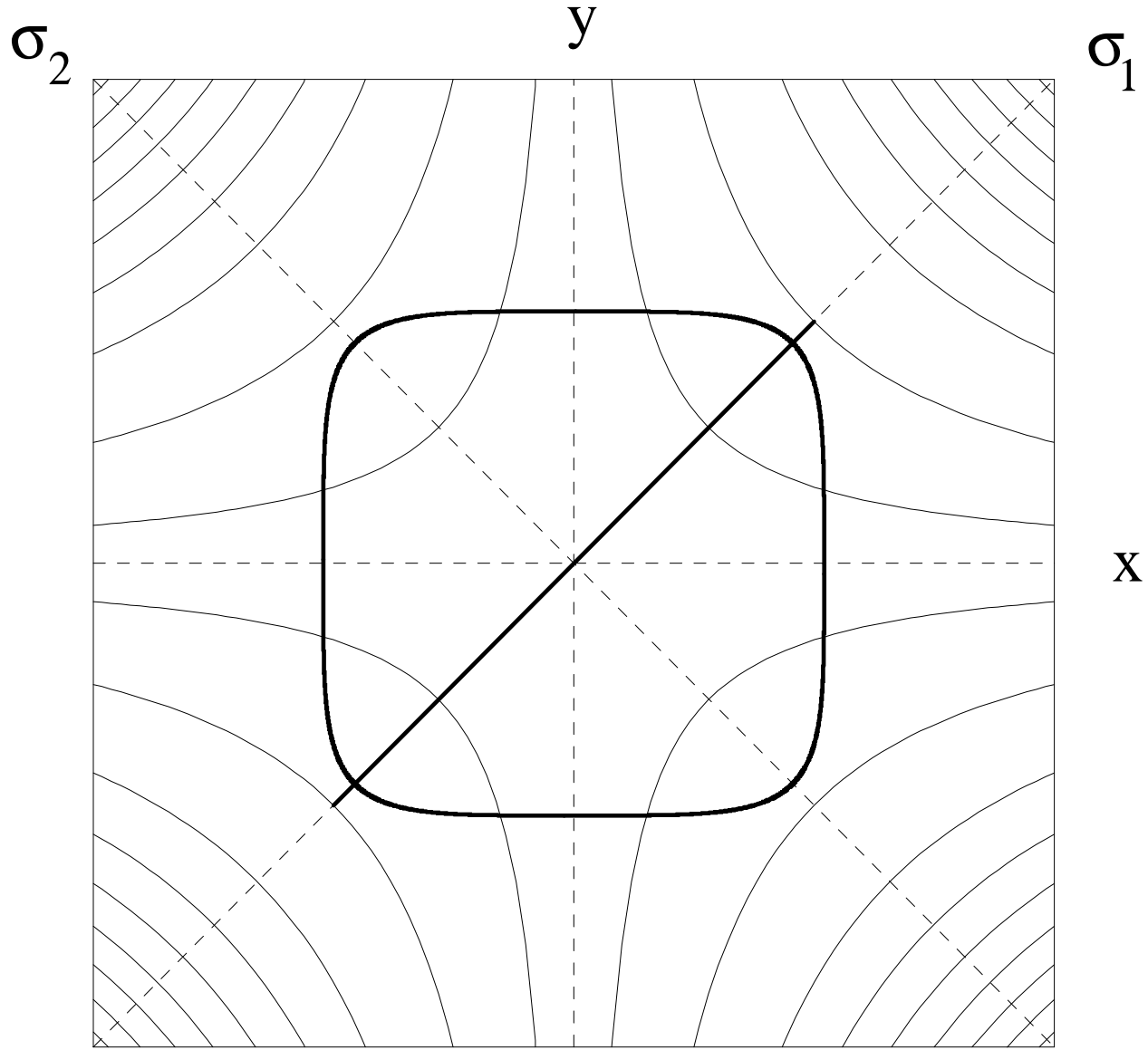
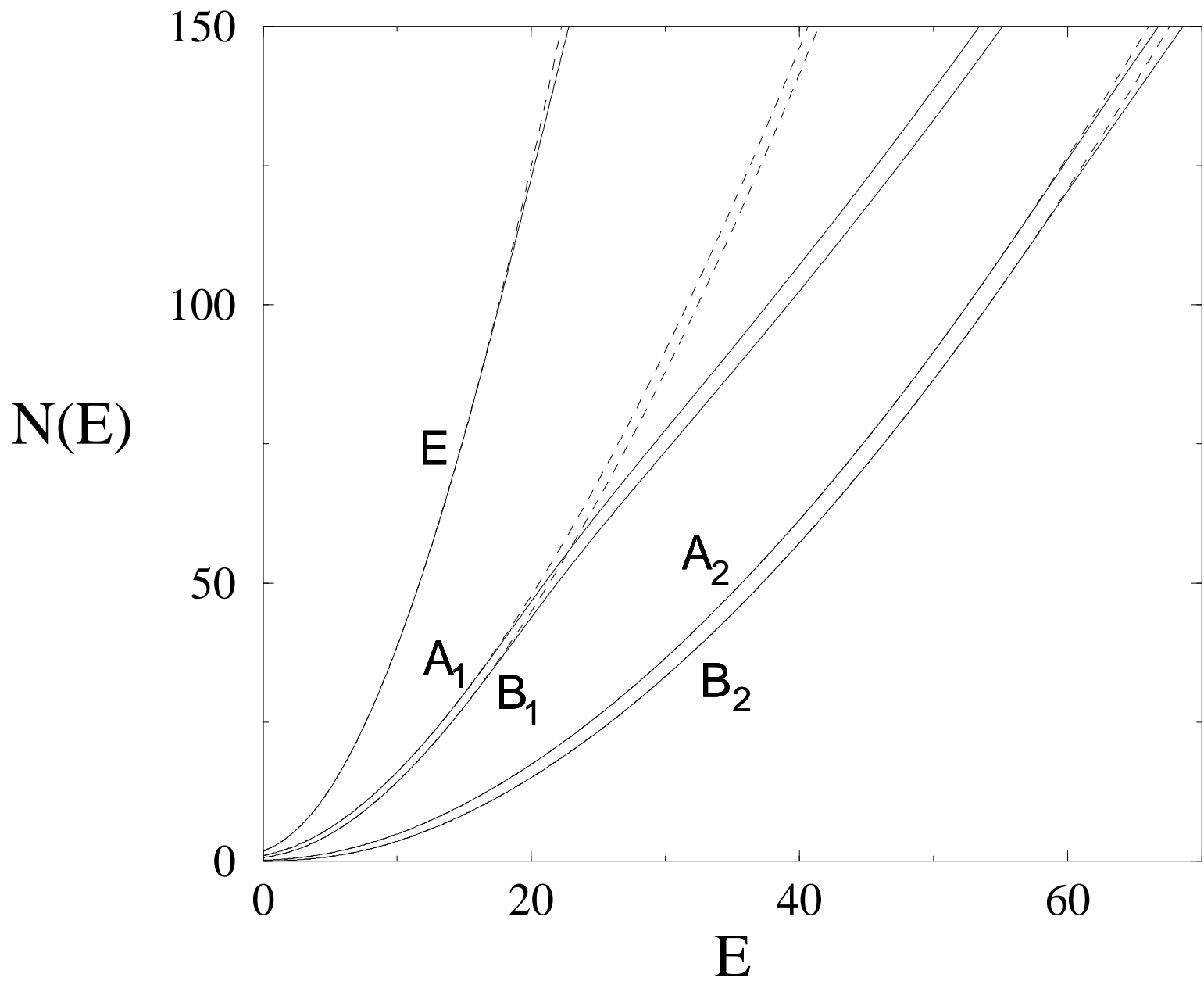
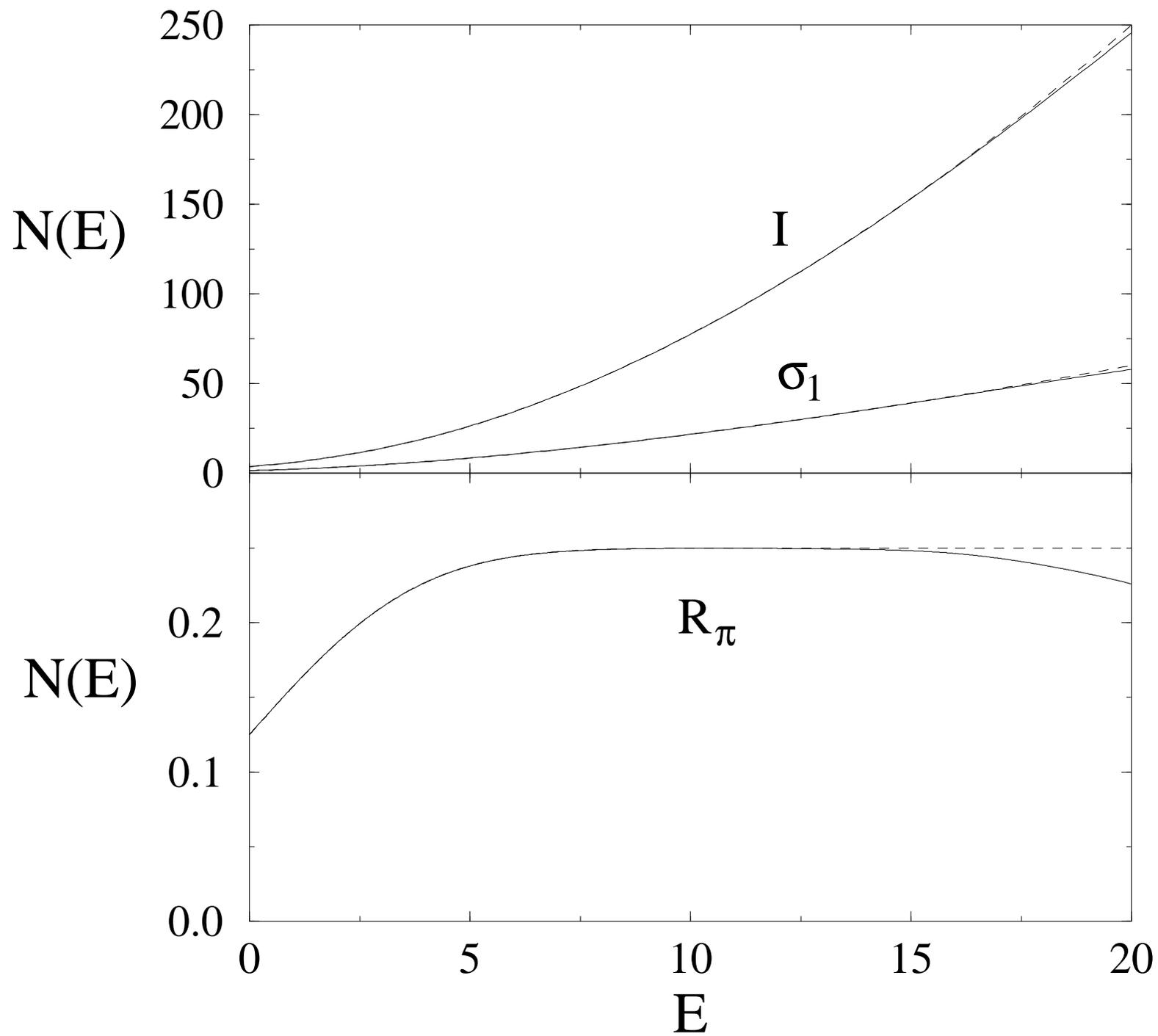
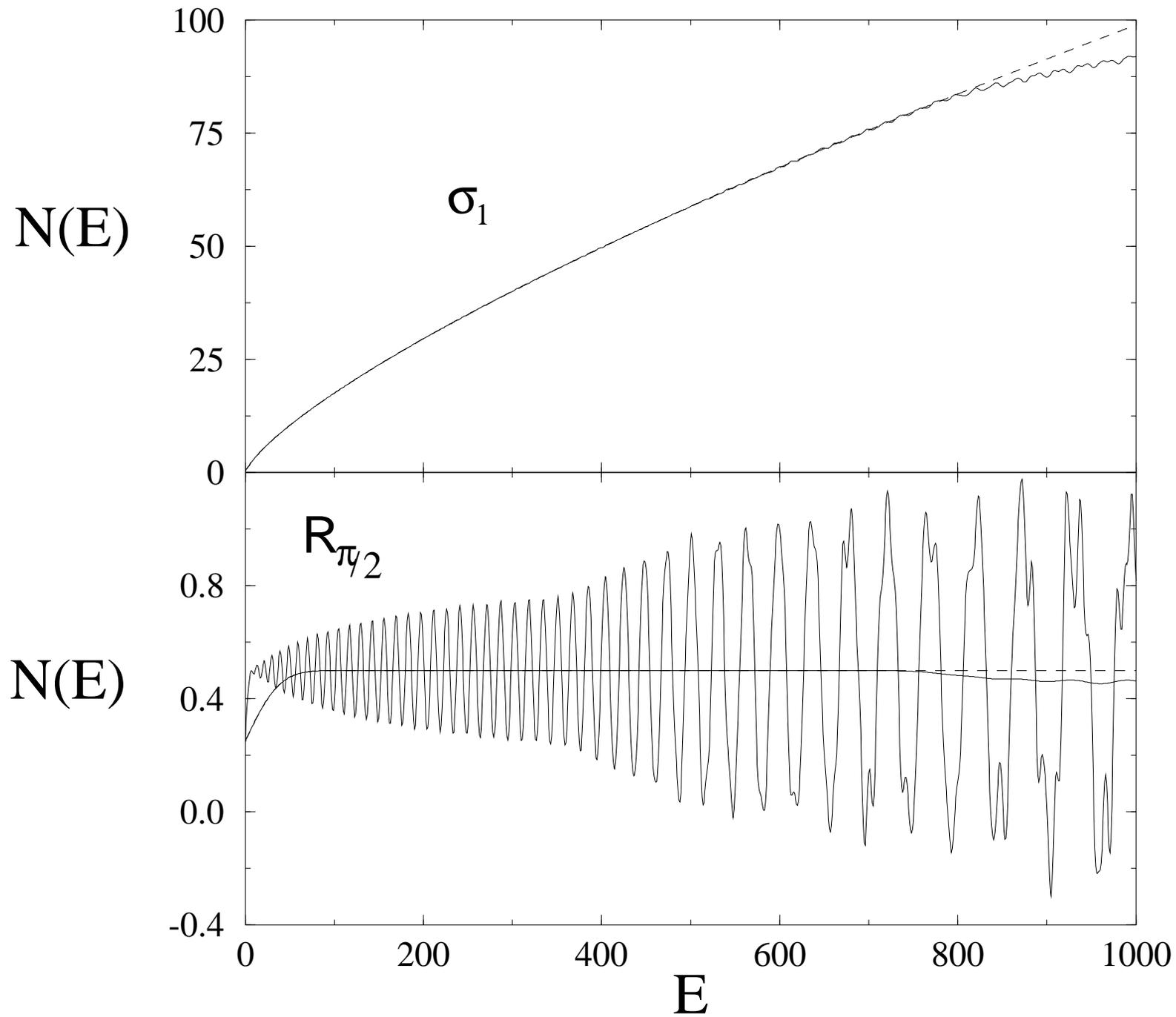
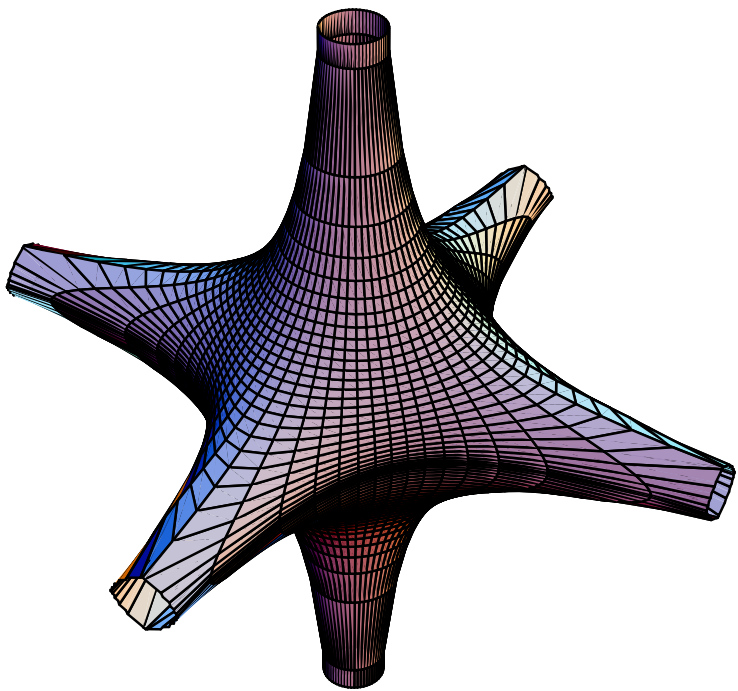
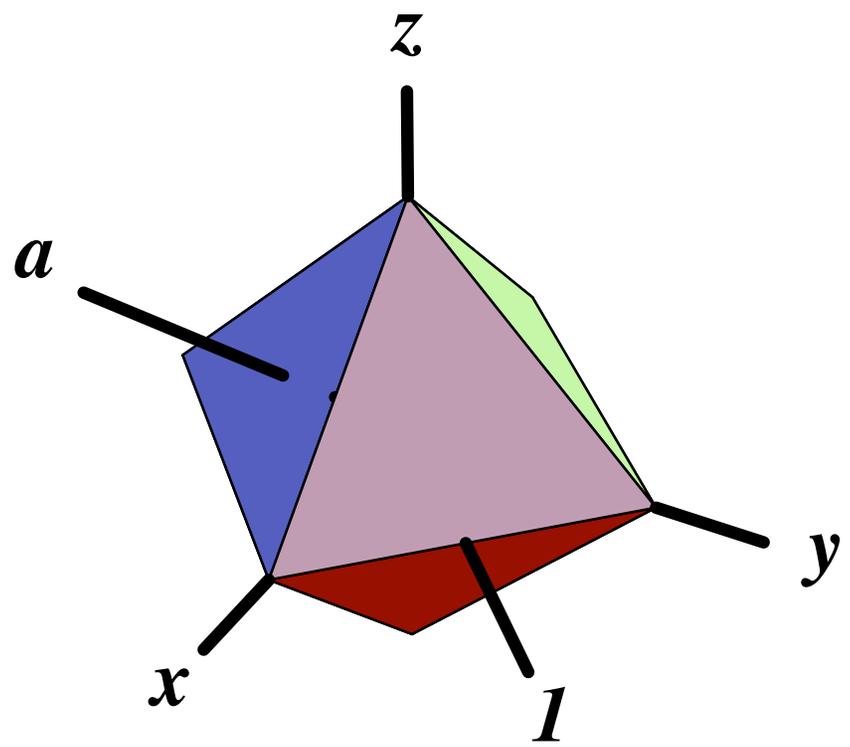
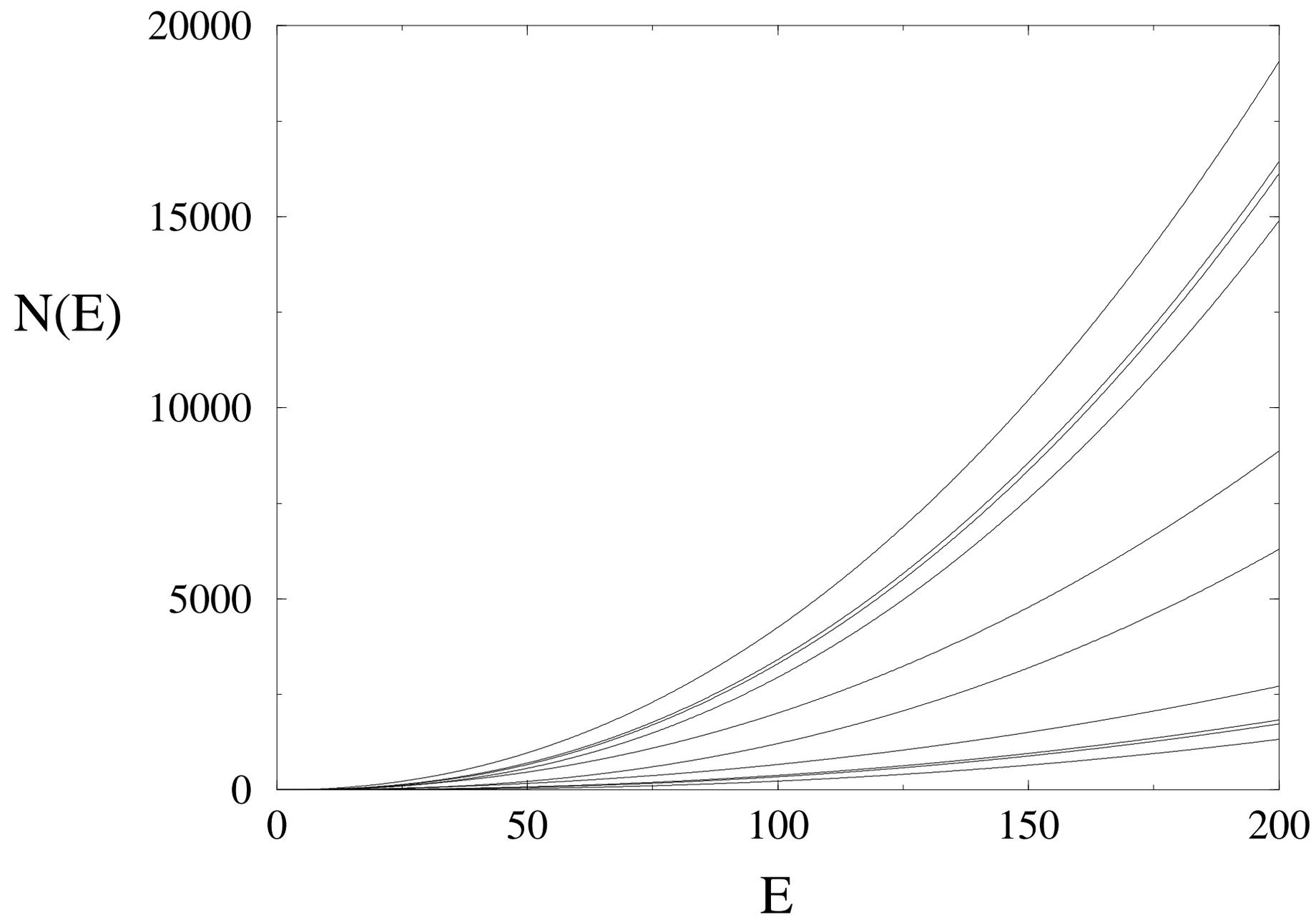
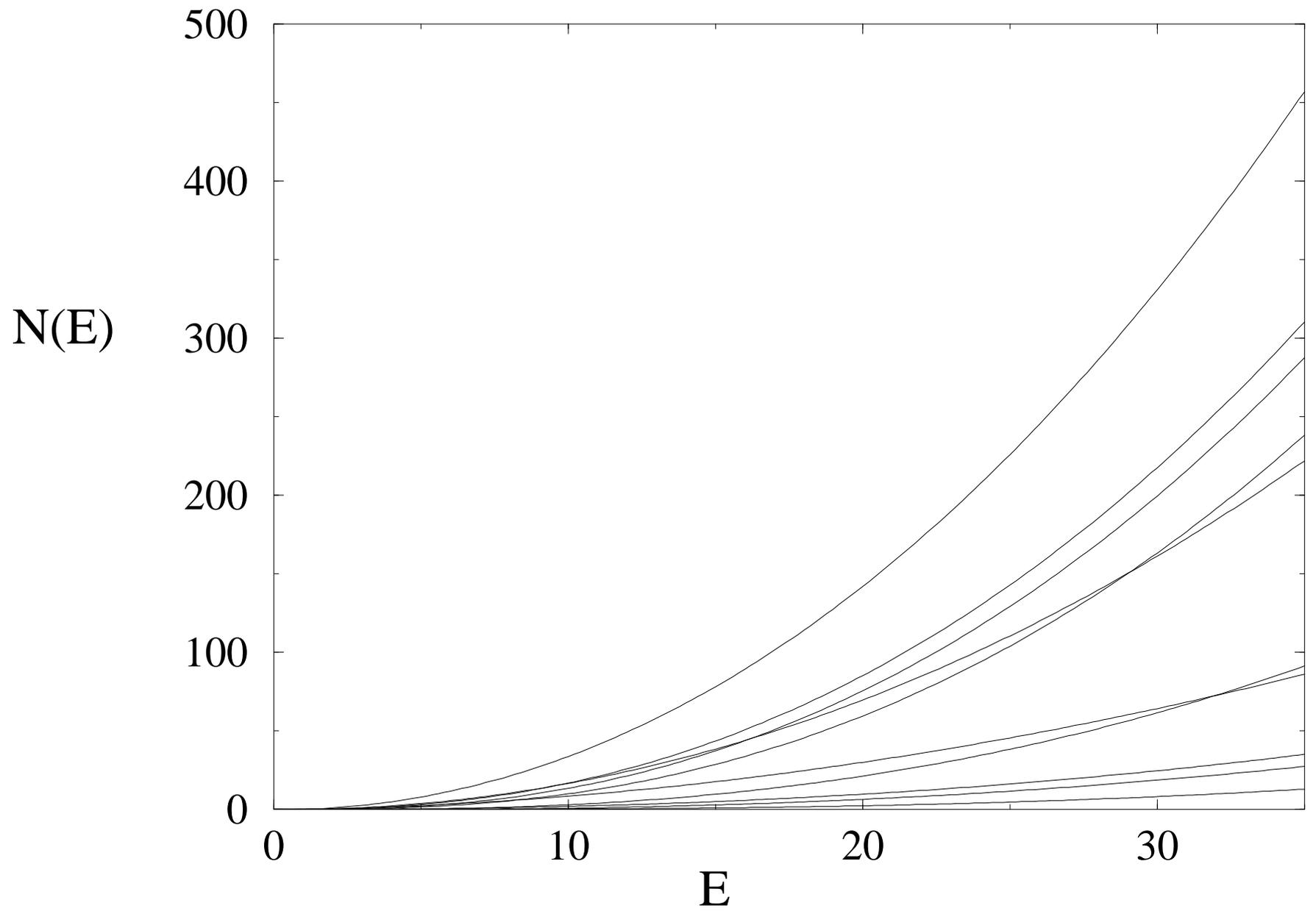
\end{document}